\documentclass[twocolumn,aip,jcp,reprint,superscriptaddress]{revtex4-1}
\pdfoutput=1

\usepackage{graphicx,color}
\usepackage{amsmath,amsfonts,amssymb}
\usepackage{dcolumn}
\usepackage{lmodern}
\usepackage{comment}
\usepackage[normalem]{ulem}
\usepackage{cancel}

\newcommand{\fig}{Fig.}

\newcommand{\figref}[1]{\fig~\ref{#1}}

\renewcommand{\eqref}[1]{equation~(\ref{#1})}
\newcommand{\Eqref}[1]{Equation~(\ref{#1})}

\newcommand{\add}[1]{#1}
\newcommand{\addr}[1]{\textcolor{blue}{#1}}
\newcommand{\rem}[1]{}%\textcolor{red}{\sout{#1}}}
\newcommand{\stkout}[1]{\ifmmode\text{\sout{\ensuremath{#1}}}\else\sout{#1}\fi}
\newcommand{\e}[1]{\text{ e}^{#1}}

\newcommand{\bfy}{{\bf y}}

\newcommand{\bfR}{{\bf R}}

\newcommand{\bfF}{{\bf F}}
\newcommand{\bfP}{{\bf P}}

\newcommand{\bfV}{{\bf V''}}

\newcommand{\bfVtilde}{{\bf \tilde{V}''}}
\newcommand{\bfzero}{{\bf 0}}
\newcommand{\bfIdent}{{\bf 1}}

\newcommand{\bfyo}{{\bf Y_{\bot}}}
\newcommand{\bfyoo}{{\bf Y_{\bot,0}}}
\newcommand{\bomega}{{\mbox{\boldmath$\Omega$}}}
\def\mem#1#2#3{  \left\langle #1 \left\vert  #2 \right\vert #3 
\right\rangle   }

\begin{document}

\title{Rate constants from instanton theory via a microcanonical approach}
%Microcanonical Instanton Rate Constants}

\author{Sean R. McConnell}
\affiliation{Institute for Theoretical Chemistry, University of Stuttgart, Pfaffenwaldring 55, 70569 Stuttgart,Germany, kaestner@theochem.uni-stuttgart.de}

\author{Andreas L\"{o}hle}
\affiliation{Institute for Theoretical Chemistry, University of Stuttgart, Pfaffenwaldring 55, 70569 Stuttgart,Germany, kaestner@theochem.uni-stuttgart.de}

\author{Johannes K\"{a}stner}
\affiliation{Institute for Theoretical Chemistry, University of Stuttgart,
  Pfaffenwaldring 55, 70569 Stuttgart,Germany, kaestner@theochem.uni-stuttgart.de}

\date{\today}

\begin{abstract}
  Microcanonical instanton theory offers the promise of providing rate constants
  for chemical reactions including quantum tunneling of atoms over the whole
  temperature range. We discuss different rate expressions, which require the
  calculation of stability parameters of the instantons. The traditional way of
  obtaining these stability parameters is shown to be numerically unstable in
  practical applications. We provide three alternative algorithms to obtain
  such stability parameters for non-separable systems, i.e., systems in which
  the vibrational modes perpendicular to the instanton path couple to movement
  along the path. We show the applicability of our algorithms on two
  molecular systems: H$_2$ + OH $\rightarrow$ H$_2$O + H using a fitted
  potential energy surface and HNCO + H $\rightarrow$ NH$_2$CO using a
  potential obtained on-the-fly from density functional calculations.
\end{abstract}

\pacs{}
%\keywords{}

\maketitle

%%%%%%%%%%%%%%%%%%%%%%%%%%%%%%%%%%%%%%%%%%%%%%%%%%%%%%%%%%%%%%%%%%%%%%%%%%
\section{Introduction}
%%%%%%%%%%%%%%%%%%%%%%%%%%%%%%%%%%%%%%%%%%%%%%%%%%%%%%%%%%%%%%%%%%%%%%%%%%

The calculation of reaction rates is a longstanding challenge in computational
chemistry.\cite{eyr31,wig32,kra40,han90,pol05} At low temperatures, quantum
tunneling of atoms must be taken into account.\cite{mei16} Instanton theory\rem{,
combined with the semiclassical approximation,} is emerging as a promising and
frequently used method to calculate tunneling rates in moderately-sized
chemical reactions. It originated in the 1960s and 1970s in 
somewhat different formulations.\cite{lan67,lan69,mil75,col77,cal77,gil77,aff81,
col88,han90,ben94,mes95,alt11} It is mostly used with a canonical ensemble, which assumes
thermal equilibration at each stage of the reaction and provides thermal rate
constants. However, even in its early stages a microcanonical formulation, with
rate constants depending on the energy, was provided.\cite{mil75} This turned
out to be rarely used. More recently, the connection between instanton theory and
Ring-polymer molecular dynamics was shown\cite{ric09,ric11} and the rate
expressions were derived from first principles.\cite{ric16} While the location
of an instanton path, an unstable periodic orbit, used to be a daunting and
numerically unstable procedure, recently improved algorithms to search for
instantons were proposed. The problem of finding an instanton path can be
turned into a saddle-point search problem\cite{arn07,jon10,ein11} for which
quantum chemistry has a rich variety of methods at hand. It turned out that a
simple modification of a truncated Newton search converges very fast and
is stable even for somewhat noisy gradients and Hessians of the potential
energy.\cite{rom11,rom11b} This allowed the location of instantons in systems
with up to 78 active atoms.\cite{rom12} Meanwhile, the canonical version of
semiclassical instanton theory is frequently used to calculate thermal rate
constants.\cite{cha75,mil94,mil95,mil97,sie99, sme03,qia07,and09,
  gou10a,gou11,gou11b, rom11,gou10,jon10,mei11,gou11a,ein11,rom12,
  kry12,kae13,alv14,kry14,mei16a,alv16,son16,lam16,lam17}

However, for bimolecular reactions it is often desirable to assume a canonical
ensemble only for the separated reactant states but not during any stage of
the reaction. Specifically, many bimolecular reactions exhibit a pre-reactive
energy minimum, a weakly bound Van-der-Waals complex. At low pressure, such a
complex does not thermally equilibrate and either proceeds over the transition
state or decays again. This limits the applicability of canonical instanton
theory.\cite{and09,mei16a} A microcanonical formulation allows the use of the
reactant's thermal distribution to calculate thermal rate constants without
assuming thermalization in a pre-reactive minimum.

Moreover, canonical instanton theory is only applicable up to a crossover
temperature $T_\text{c}$, the temperature where the instanton path collapses
to a point. While different approaches to extend the formulation above
$T_\text{c}$ have been suggested\add{,\cite{han88,kry11,kry13,zha14}}
a microcanonical formulation provides thermal rate constants at all temperatures
naturally.\cite{ric16a}

Even though algorithms to calculate microcanonical instanton rate constants were
proposed decades ago\cite{mil75,ric16,ric16a} they were
rarely used for real chemical reactions in which the vibrational modes are not
separable from each other and, most importantly, from the transition mode. The
reason for this is that these approaches lacked numerical stability. In this paper we
propose different algorithms to calculate microcanonical instanton rate
constants for non-separable systems. We compare results to canonical instanton
theory, as well as to results from exact quantum dynamics.

The paper is organized as follows: First, we briefly review the theory of
microcanonical rate constants and ways to derive the cumulative reaction
probability $P(E)$ using instanton theory and the semiclassical
approximation. The resulting rate expressions require the calculation of
stability parameters. Besides the traditional approach of solving the
stability matrix differential equation, we provide three alternative,
numerically more stable, approaches to calculate the stability parameters. In
the applications section we apply these to the two test cases H$_2$ + OH
$\rightarrow$ H$_2$O + H and HNCO + H $\rightarrow$ NH$_2$CO. Finally we
discuss advantages and disadvantages of our newly proposed approaches and of
microcanonical instanton theory in general.

%%%%%%%%%%%%%%%%%%%%%%%%%%%%%%%%%%%%%%%%%%%%%%%%%%%%%%%%%%%%%%%%%%%%%%%%%%
\section{Theory}
%%%%%%%%%%%%%%%%%%%%%%%%%%%%%%%%%%%%%%%%%%%%%%%%%%%%%%%%%%%%%%%%%%%%%%%%%%

%%%%%%%%%%%%%%%%%%%%%%%%%%%%%%%%%%%%%%%%%%%%%%%%%%%%%%%%%%%%%%%%%%%%%%%%%%
\subsection{Microcanonical Reaction Rate Constants}

In order to describe bimolecular reactions to their full extent one would have
to solve the Schr\"odinger equation with proper scattering boundary conditions
in order to obtain the S-matrix which contains all the necessary information
to calculate the state-to-state differential and integral cross sections. By
averaging over all cross sections one obtains the so called cumulative
reaction probability\cite{mil75}.
%EQUATION 1
\begin{align}
  P(E) = \sum_J \left(2 J + 1 \right)\sum_{n_p, n_r} \left|\mathbf{S}_{n_p, n_r} (E, J) )\right|^2 
\end{align}
whereby $n_p$ and $n_p$ denote the quantum numbers of the product and reactant state, $J$ the total angular momentum and $E$ the total energy of the system. The microcanonical rate constant $k(E)$ can then be calculated as follows \cite{mil91} 
%EQUATION 2
\begin{align}  
k(E) = \frac{1}{2 \pi \hbar} \frac{P(E)}{\Gamma_r(E)}
\end{align}
where $\Gamma_r(E)$ is density of states in the reactant state. The
canonical rate constant $k(T)$ is obtained via a Laplace transform of $P(E)$
%EQUATION 3
\begin{align}
k(T) = \frac{1}{2 \pi \hbar Q_{\text{RS}}} \int_{-\infty}^{\infty} P(E) \exp(-\beta E) dE
\end{align}
divided by the canonical partition function $Q_{\text{RS}}$ of
the reactant state per unit volume. Here $\beta$ is the inverse temperature, $\beta=1/(k_\text{B}T)$. 
%\newline

Since in most cases one is only interested in obtaining the chemical reaction rate rather than detailed information about all state to state interactions (which would be provided by a full scattering calculation) an efficient way to obtain $P(E)$ directly is the use of the quantum flux-flux autocorrelation  formalism \cite{mil98a} which gives an exact expression for $P(E)$
%EQUATION 4
\begin{align}
  P(E) = 2 \pi \hbar \ \text{tr}(\delta(E-\hat{H}) \hat{F} \hat{P}_r) \label{trace1}
\end{align}
where $\delta(E-\hat{H})$ is the density operator in the microcanonical ensemble. $\hat{F}$ is the quantum mechanical analogue of the classical flux function which counts the number of elementary reactions from reactant to product and is given by
%EQUATION 5
\begin{align}
  \hat{F} = \frac{\text{i}}{\hbar} \left[\hat{H}, \hat{\theta} (s) \right]
\end{align}
where $\hat{\theta}$ is the Heaviside step function and $s$ denotes a function that is negative on the reactant side of the dividing surface and positive on the product side. The projection operator $\hat{P}_r$ is given by the time evolved Heaviside function in the limit of $t \to \infty$
%EQUATION 6/7
\begin{align}
  \hat{P}_r &= \lim_{t \to \infty }  e^{\frac{\text{i}}{\hbar} \hat{H} t} \hat{\theta}(s) e^{-\frac{\text{i}}{\hbar} \hat{H} t} \\
            &= \int_0^{\infty}  e^{\frac{\text{i}}{\hbar} \hat{H} t} \hat{F}  e^{-\frac{\text{i}}{\hbar} \hat{H} t} dt \label{time_int}
\end{align} 
and can be written as the time integral of the time evolved flux operator. It
describes the probability that the trajectory remains on the the product side
as $t$ approaches infinity. After some manipulations of \eqref{trace1} one arrives at the final expression for the cumulative reaction probability \cite{mil83} 
%EQUATION 8
\begin{align}
  P(E) = 2 \pi^2 \hbar^2 \text{tr}\left(\delta(E-\hat{H}) \hat{F} \delta(E-\hat{H}) \hat{F} \right) \label{final}
\end{align}
Over the years there have been several methods \cite{mil98a, mil91, man93a, thi85} proposed to evaluate \eqref{final} which vary significantly in terms of accuracy and computational effort. However, for large systems, a semi-classical approximation of \eqref{final} remains the method of choice. In this paper we use a formulation of instanton theory to evaluate $P(E)$ which has been recently proposed by Richardson based on the previous works of Miller in which the evaluation of \eqref{final} is reduced to finding closed orbits in imaginary time  and the calculation of its stability parameters $u_i$.\cite{gut71}

%%%%%%%%%%%%%%%%%%%%%%%%%%%%%%%%%%%%%%%%%%%%%%%%%%%%%%%%%%%%%%%%%%%%%%%%%%
\subsection{The Cumulative Reaction Probability $P(E)$}

We consider scattering problems, i.e. situations in which the reactant
state is unbound and can adopt a continuum of energy values. This
corresponds to a bimolecular reaction. The thermal rate constant
$k(\beta)$ can be obtained from the cumulative reaction probability
$P(E)$ via
%EQUATION 9
\begin{equation}
  k(\beta) Q_\mathrm{RS}(\beta)=
  \frac{1}{2\pi\hbar}\int_{E_\text{RS}}^\infty
    P(E)\exp(-\beta E) dE
    \label{eq:mic:can}
\end{equation}
where $E_\text{RS}$ is the energy of the reactant state in its
vibrational ground state, $Q_\text{RS}$ is the partition function of
the reactant state per unit volume and $\beta$ is the inverse
temperature, $\beta=1/(k_\text{B}T)$. Atomic units with
$\hbar=m_\text{e}=4\pi\epsilon_0=1,\,c=1/\alpha$ will be used from now on.

\begin{figure}%[h!]
\begin{center}
  \includegraphics[width=8cm]{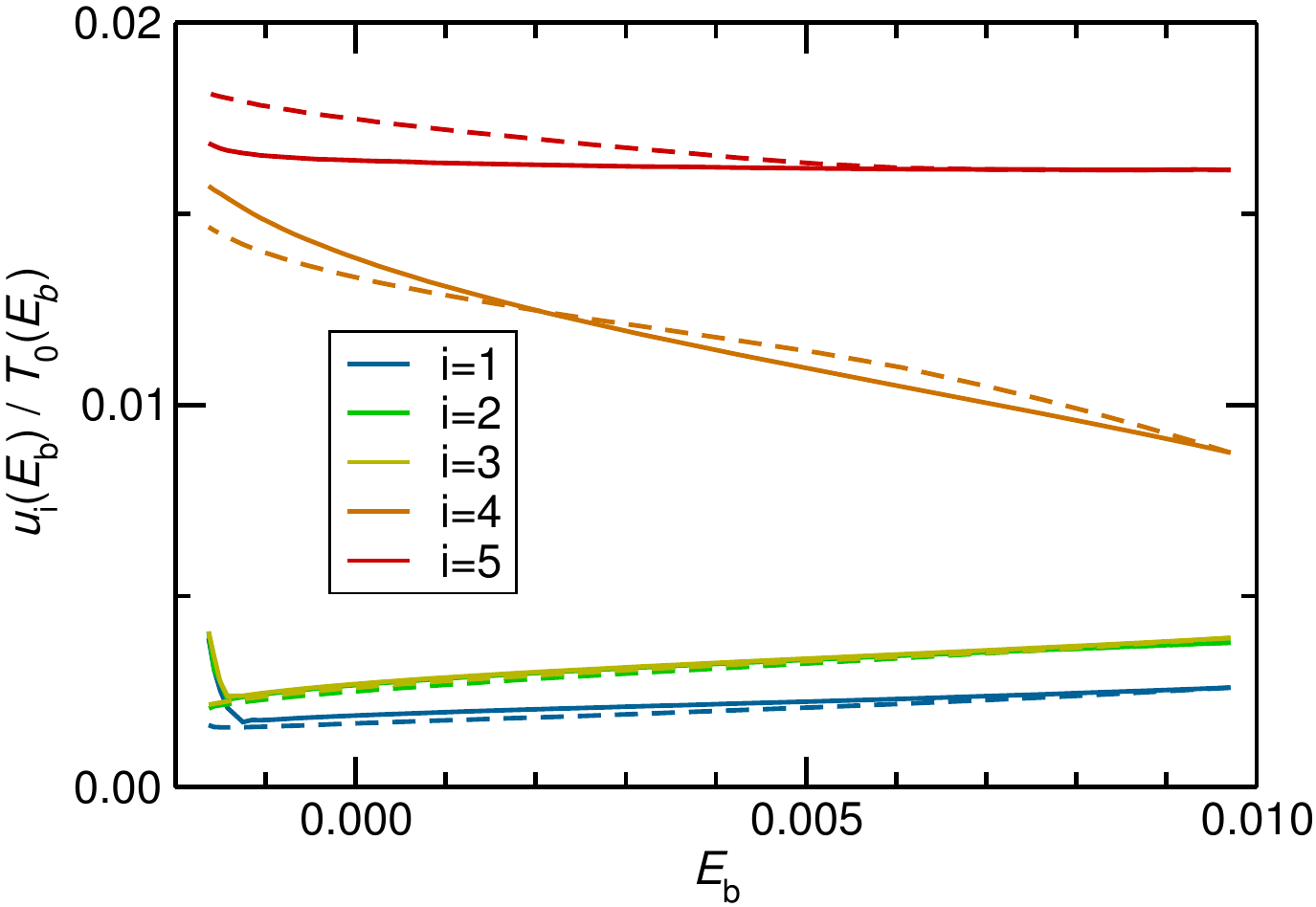}
  \caption{The stability parameters $u_i(E)$ for the reaction H$_2$+OH
    $\rightarrow$ H$_2$O + H $\rightarrow$ H$_2$O + H discussed in the results section. All values in
    atomic units. Solid lines were obtained by solving the stability matrix
    differential equation, dashed lines by eigenvalue
    tracing. 
%    \sm{We should note (probably in the discussion), that too few images will render e-val 
%    tracing useless. A loss of coherence between neighbouring, rotated, reduced Hessians 
%    means that a true avoided crossing may be missed or vice-versa. There may be some systems 
%    in which a prohibitively large number of images are required to maintain this coherence.}
    \label{fig:mic:stabpar}
  }
\end{center}
\end{figure}

In one dimension, $P(E)$ can be obtained by a variety of methods,
including a direct numerical solution of Schr\"odinger's
equation. Instanton theory provides an expression for $P(E)$ for a
system with $D$ vibrational degrees of freedom:\cite{mil75}
%EQUATION 10
\begin{align}
  P(E)=&\sum_{k=1}^\infty(-1)^{k-1}\exp(-kS_0(E))\times\nonumber\\
   &\prod_{i=1}^{D-1}\frac{1}{2\sinh(ku_i(E)/2)} 
  \label{eq:mic:kseries}
\end{align}
Where $S_0$ is the shortened action and $u_i(E)$ are the stability
parameters of the instanton path. \add{Their dependence on $E$ for a specific model system is displayed in
\figref{fig:mic:stabpar}.}
$D$ is the number of vibrational
degrees of freedom of the system. The shortened action is
%EQUATION 11
\begin{equation}\label{eq:short_action}
  S_0(E_b)=\int_0^{T_0}
  \left|\frac{d\bfy(\tau)}{d\tau}\right|^2d\tau=\sqrt{8}\int_{r_a}^{r_b} \sqrt{E(r)-E_b}dr
\end{equation}
with the integration being done along the instanton path \rem{$\bfy(t)$} \add{$\bfy(\tau)$} in
mass-weighted coordinates $\bfy$. \add{The integral on the left in \eqref{eq:short_action} 
is performed in complex time $t\to i\tau$}. In real-space the integration can be
done via the arc length $r$ between the turning points $r_a$ and $r_b$
with $E(r_a)=E(r_b)=E_b$. The instanton optimization provides a
tunneling energy $E_b$ for a given $T_0$. 

With the hyperbolic sine expressed as its series expansion,
\eqref{eq:mic:kseries} results in:
%EQUATION 13
\begin{equation}
  \frac{1}{2\sinh(ku_i(E)/2)}=\sum_{n_i=0}^\infty
  \exp\left[-k(n_i+\tfrac 1 2)u_i(E)\right]
\end{equation}
from which we arrive at
%EQUATION 14
\begin{equation}
  P(E)=\sum_n\sum_{k=1}^\infty   (-1)^{k-1}\e{-k\left[S_0+ 
      \sum_{i=1}^{D-1}(n_i+\tfrac 1 2)u_i(E)\right]}
\end{equation}
where the short-hand notation
%EQUATION 15
\begin{equation}
  \sum_n=\sum_{n_1=0}^\infty \sum_{n_2=0}^\infty \ldots
  \sum_{n_{D-1}=0}^\infty
\end{equation}
is used.  The sum over $k$ can be interpreted as a geometric
series\cite{mil75} and can be calculated explicitly:
%EQUATION 16
\begin{equation}
  P(E)=\sum_n \left\{1+\exp\left[S_0(E)+ 
      \sum_{i=1}^{D-1}(n_i+\tfrac 1 2)u_i(E)\right]\right\}^{-1}
  \label{eq:mic:noshift}
\end{equation}
Applicability of the equation is limited in multidimensional systems,
since we need $P(E)$ for energies above the energy of the reactant
including the zero-point vibrational energy (ZPE)
$E_\text{RS,ZPE}$. $S_0(E)$, however, is only available for
$E<E_\text{TS}$, i.e., the energy of the saddle point without ZPE. In
general, for multidimensional systems, $E_\text{TS}$ is often smaller
than $E_\text{RS,ZPE}$. However, at least for systems where the
vibrational modes are separable for the whole instanton path, the
physical background makes it clear that $P(E)$ should be independent
of the vibrational frequencies perpendicular to the transition mode,
and, thus, the ZPE.\cite{ric16a}

\begin{figure}[h!]
\begin{center}
  \includegraphics[width=8cm]{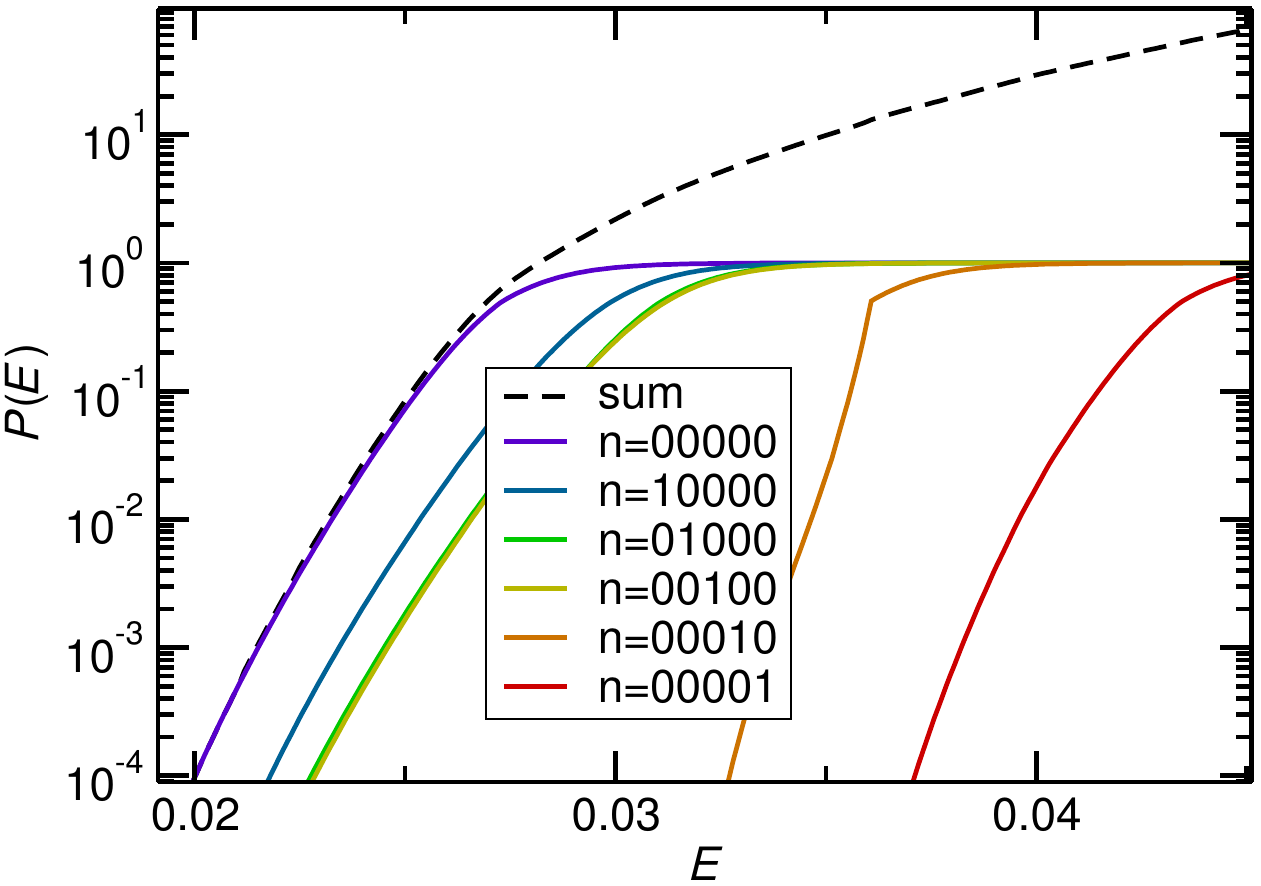}
  \caption{The individual terms of \eqref{eq:mic:2} and their sum,
    $P(E)$, with the stability parameters calculated by eigenvalue
    tracing. All values in atomic units. It is notable that
the energy-dependence of the contribution with $u_4$ differs from the
others. This is the stability parameter which depends strongly on $E$,
see \figref{fig:mic:stabpar}.
    \label{fig:mic:p:stabpar}
  }
\end{center}
\end{figure}

A way to circumvent that dilemma which is also applicable to
non-separable systems was proposed a long time
ago,\cite{mil75,cha75} \add{resulting in} 
%\begin{equation}
%remove 
%\end{equation}
%\begin{equation}
%remove 
%\end{equation}
\begin{equation}
%\begin{multline}
 E_n=\sum_{i=1}^{D-1}(n_i+\tfrac 1 2)\frac{u_i(E-E_n)}{T_0(E-E_n)}
  \label{eq:mic:stabpar}
%\end{multline}
\end{equation}
and
%EQUATION 20
\begin{equation}
  P(E)=\sum_n\frac{1}{1+\exp[S_0(E-E_n)]}.
  \label{eq:mic:2}
\end{equation}
With that transformation, $S_0$ and $u_i$ \rem{(via $\omega_i$)} are
required in the energy range where they can be easily
calculated. \Eqref{eq:mic:stabpar} needs to be solved iteratively. The
individual terms of \eqref{eq:mic:2} and the sum are displayed in
\figref{fig:mic:p:stabpar}. 

%The individual terms of equation (11) and their sum, $P(E)$ for the
%reaction H$_2$+OH discussed in the results section.

At the energy $E_\text{TS}=E_\text{c}-E_n$ the instanton collapses to one point,
$S_0=0$ and $u_i=T_{0,\text{c}}\omega_{i,\text{TS}}$ with $T_{0,\text{c}}$
being the inverse critical temperature
$T_{0,\text{c}}=\beta_c=2\pi/\bar\omega_{\text{TS}}$ and
$\bar\omega_{\text{TS}}$ is the absolute value of the imaginary frequency
at the transition state.  Above that energy, $P(E)$ is not accessible any more
from instanton theory. Instead, we follow a previous suggestion\cite{ric16a}
and use the exact transmission coefficient of a parabolic barrier with the
barrier frequency $\bar\omega_{\text{TS}}$:\cite{bel80}
%EQUATION 21
\begin{equation}
P_{\text{parabolic}\add{,n}}(E)=\frac{1}{1+\exp[2\pi(E_\text{TS}+E_n-E)/\bar\omega_{\text{TS}}]}
\end{equation}
Almost indistinguishable results are obtained when using the transmission
coefficient of a symmetric Eckart barrier with the same barrier height and
frequency as the real barrier. This approach may lead to a kink in the
contribution to $P(E)$, as visible, for example, for $n=(00010)$ in
\figref{fig:mic:p:stabpar} (orange line), which will be averaged out,
however, when calculating $k(T)$.

As discussed below, for non-separable systems at low temperature, the
calculation of the individual stability parameters $u_i$ can be
numerically unstable. In these cases it is often still possible to
calculate the term as a multi-dimensional integral\citep{ben92,ben92a,kry11}
%EQUATION 22
\begin{multline}\label{eq:sigma_def}
  \exp(-\sigma)=\prod_{i=1}^{D-1}\frac{1}{2\sinh(u_i(E)/2)}=\prod_{i=1}^{D-1}\int d\bfyo^{(i)}(\tau)\times \\
       {\exp{\left[-\int_{0}^{T_0}\bfyo^{(i)}(\tau)^T\left(-\frac{1}{2}
           \frac{d^2}{d\tau^2}+\frac{1}{2} 
           \bfV(\tau)\right)\bfyo^{(i)}(\tau)d\tau\right]}}  
\end{multline}
%  \exp(-\sigma)=\prod_{i=1}^{D-1}\frac{1}{2\sinh(u_i(E)/2)}=\oint D\left[\bfyo(\tau)\right]\times \\
%  \exp{\left[-\int_{0}^{T_0}\bfyo(\tau)^T\left(\frac{1}{2}
%           \frac{d^2}{d\tau^2}+\frac{1}{2} 
%           \bfV(\tau)\right)\bfyo(\tau)d\tau\right]}
%where $\sigma$ can be expressed as a multi-dimensional integral\citep{Benderskii1992,Benderskii1992_2,kry11}
%\begin{equation}\label{eq:sigma_def2}
%  \sigma=\mathrm{tr}\left(\int_{0}^{T_0}\bfyo(\tau)^T\cdot\left(\frac{1}{2}
%           \frac{d^2}{d\tau^2}+\frac{1}{2} 
%           \bfV(\tau)\right)\cdot\bfyo(\tau)d\tau\right).
%\end{equation}
Here, the matrix $\bfyo(\tau)$ is a co-moving basis containing all vibrational
modes orthogonal to the instanton path\add{, $\bfyo^{(i)}(\tau)$ is the \emph{i}-th column 
vector of $\bfyo(\tau)$} and $\bfV$ is the matrix of second
derivatives of the potential energy with respect to mass-weighted Cartesian
coordinates. $P(E)$ can then be approximated by
%EQUATION 23
\begin{equation}
  P(E)={\sum_n}\frac{1}{1+\exp[S_0(E{-E_{\text{vib},n}}-\sigma/T_0)]}
  \label{eq:mic:psigma}
\end{equation}
\add{with
\begin{equation}
  E_{\text{vib},n}=\sum_{i=1}^{D-1}\omega_{i,\text{TS}}\ n_i
\end{equation}
where $\sigma/T_0$ approximates the zero-point vibrational energy and
$E_{\text{vib},n}$ covers vibrational excitations. In many cases, the first
term of \eqref{eq:mic:psigma}, the one with $E_{\text{vib},n}=0$, dominates.}
A similar approach has been suggested recently.\cite{ric16a}

In a numerical implementation, instantons are optimized for a given set of
oscillation times $T_0$ (or temperatures $\beta=T_0/\hbar$). These provide
sets of $S_0$, $E_b$, $u_i(E)$ or $\sigma(E)$. The properties
required to calculate $P(E)$ for a given energy $E$ are interpolated. When
individual stability parameters are used, $u_i(E)/T_0(E)$ is linearly
interpolated to iteratively solve \eqref{eq:mic:stabpar}. Then $S_0$ is
linearly interpolated to obtain $P(E)$ via \eqref{eq:mic:2}. If
\eqref{eq:mic:psigma} is used, then $S_0$ is linearly interpolated between
the two neighboring occurrences of $E=E_b+\sigma/T_0$.

At low energy and for large vibrational frequencies, the sum over $n$
in \eqref{eq:mic:2} can be truncated after a few terms, possibly even
after the first term. At high energies, especially when
$E>E_\text{TS,ZPE}$, many terms must be included.  At high enough
energies, however, $S_0=0$ and can be assumed independent of $E_n$. The
quantization of vibrational energy levels can therefore be neglected, which results
in
%EQUATION 24
\begin{equation}
  P(E)=\frac{(E-E_\text{TS})^{D-1}}{(D-1)!}  \prod_{i=1}^{D-1}
  (\omega_{i,\text{TS}})^{-1} \quad \text{ for } E\gg E_\text{TS}.
  \label{eq:qts:pofe:highE}
\end{equation}
We use \eqref{eq:qts:pofe:highE} for energies above $E_\text{TS,ZPE}+10\times\omega_\text{TS,min}$, with
$\omega_\text{TS,min}$ being the smallest vibrational frequency at the
transition state perpendicular to the transition mode. This ensures
that the fist 10 quanta of the vibrations are taken into account
explicitly and the continuous expression is used above. 

Up to now, we have only discussed the treatment of vibrational
levels. We consider the rotational motion to be separable from the
internal motion. Our $P(E)$ is essentially $P(E,J)$ for $J=0$. In the
$J$-shifting approximation\cite{tak52,bow91} the rotation-dependence
of $P(E,J)$ is taken out of the integral in \eqref{eq:mic:can}:
%EQUATION 25
\begin{multline}
  k(\beta) Q_\mathrm{RS}(\beta)=
  \frac{1}{2\pi} Q_\text{rot,TS} \times\\\int_{E_\text{RS}}^\infty
    P(E,J=0)\exp(-\beta E) dE
    \label{eq:mic:can:rot}
\end{multline}
The rotational partition function of the transition state
$Q_\text{rot,TS}$ is approximated by is classical expression. This is
generally a good approximation. The moments of inertia are obtained
from the transition state geometry.

In practice, instantons are located at pre-defined $T_0$ (or temperatures) by
sequential cooling. To cover the full temperature range, instantons need to be
located until $E_b+\sigma/T_0<E_\text{ZPE,RS}$. The thermal rate is then
obtained via
%EQUATION 26
\begin{equation}
  k(\beta)=
  \frac{1}{2\pi}\frac{Q_\text{rot,TS}}{ Q_\mathrm{RS}(\beta)}\int_{E_\text{ZPE,RS}}^\infty
    P(E)\exp(-\beta E) dE
    \label{eq:mic:can:final}
\end{equation}

%%%%%%%%%%%%%%%%%%%%%%%%%%%%%%%%%%%%%%%%%%%%%%%%%%%%%%%%%%%%%%%%%%%%%%%%%%
\subsection{The Stability Parameters $u_i$}

Using equations~(\ref{eq:mic:stabpar}) and (\ref{eq:mic:2}) or
\eqref{eq:mic:psigma} require the calculation of the stability
parameters $u_i$ or at least of their combination in  the form of $\sigma$. Methods to
calculate these have appeared in the literature. They almost exclusively
consist of integrating the stability matrix differential equation.\cite{mil75,
 gar00,bad03,car07} However, it is clear that this approach is
numerically unstable for strong coupling between the modes and/or for a
small number of images $P$ discretizing the instanton path. The number of
images must be kept small, though, to keep the computational effort at bay when
dealing with energies and its derivatives calculated on the fly. We derived
and tested several approaches to calculate $u_i(E)$ or $\sigma(E)$ and here
report on the four that proved numerically most stable in practice.

All the algorithms described here were implemented in a development version of
the open-source general-purpose geometry optimizer DL-FIND.\cite{kae09a} The
code will be made available to the scientific community in due course.

%%%%%%%%%%%%%%%%%%%%%%%%%%%%%%%%%%%%%%%%%%%%%%%%%%%%%
\subsubsection{The stability matrix differential equation\label{sec:stab_mat}}  	

The stability parameters $u_i$ are found by solving the linearized 
equations of motion for the stability matrix 
$\bfR(\tau)$\cite{gut71,mil75,kle09}
%EQUATION 27                
\begin{equation}
  \frac{d}{d\tau}\bfR(\tau)+\bfF(\tau)\bfR(\tau)=0
\label{GYeq}
\end{equation}		
where $\bfR$ is a $2D\times 2D$ matrix and
%EQUATION 28
\begin{equation}
  \bfF(\tau)=\left(\begin{array}{cc}
                     \bfzero & -\bfIdent\\
                     -\bfV (\tau) & \bfzero
                   \end{array}\right).
 \label{GYF}
\end{equation}	
The matrix $\bfV$ is the matrix of second derivatives of the potential
energies with respect to the mass-weighted coordinates of the atoms at
the point $\tau$ along the instanton path. \Eqref{GYeq} must be solved
for $\bfR(T_0)$ with the initial condition $\bfR(0)=\bfIdent$.
%Solving equation (\ref{GYeq}) for either $\bfA_1$ or $\bfA_2$
%reproduces the canonical form of the Gel'fand--Yaglom equation.
The eigenvalues of $\bfR(T_0)$ are in pairs $e^{u_i}$ and $e^{-u_i}$ for each
$i$. Besides the $D-1$ stability parameters, $\bfR(T_0)$ has two additional
eigenvalues which are unity and correspond to the movement along the path. For
a molecular system, there are additional 10 (for linear molecules) or 12
eigenvalues of $\bfR(T_0)$ which are unity and correspond to the translation
and rotation of the total system.  In practice, \eqref{GYeq} is solved with an
implicit \add{(or backward)} Euler algorithm \add{or, alternatively, a
  fourth-order Runge--Kutta approach (RK4),} by discretization using the
images of the instanton path. \add{RK4 is used in the results section unless
  noted otherwise.} Solving the stability matrix is a reliable technique when
instanton paths are short.  At lower energies, depending on the number of
images $P$, the eigenvalues which correspond to movement along the path become
indistinguishable from the $u_i$ and the algorithm becomes numerically
unstable, as can be seen in \figref{fig:mic:stabpar} by the increase of
$u_i(E)/T_0(E)$ for the lowest three stability parameters (blue, green and
yellow solid curves) at low energies. \add{Applicability of the method can be
  extended by using solely $e^{u_i}$ and ignoring $e^{-u_i}$, which becomes
  small and may become negative due to numerical noise. At too low energies,
  eigenvalues which are supposed to be used to calculate $u_i$ show non-zero
  imaginary parts. In these cases, we extrapolate by using $u_i(E)$ from the
  lowest energy for which valid $u_i$ were obtained.}

%--------------------------------------------------------------
\subsubsection{Stability parameters by eigenvalue tracing}

\addr{An approximation to equations~(\ref{GYeq}) and (\ref{GYF}) can
  be found by realizing} that for slowly-varying
  frequencies, the stability parameters $u_i(E)$\rem{, especially in their form
$\omega_i(E)=u_i(E)/T_0(E)$} can be interpreted as frequencies \add{$\omega_i(\tau)$}
perpendicular to the instanton path averaged along that path,
\begin{equation}
  u_i(E)=\int_0^{T_0}\omega_i(\tau)d\tau.
  \label{eq:uofomega}
\end{equation}

To achieve that averaging, individual vibrational frequencies need to
be traced along the instanton path and then averaged. To do this,
we first construct a reduced $(D-1)\times (D-1)$ Hessian matrix
$\bfVtilde$ at each image of the instanton. This is found by
%EQUATION 29
\begin{equation}
  \bfVtilde=\bfyo^T \bfV \bfyo,
\end{equation}
i.e., projecting the full Hessian onto a basis $\bfyo$ which contains all
modes perpendicular to the instanton path at that image and
perpendicular to the translational and rotational eigenvectors. The
tangent vector of the instanton path is provided by the eigenvector
$v_\text{tang}$ of the Hessian of the full instanton, the eigenvalue of
which is zero. The basis $\bfyo$ \emph{does not} contain the mode 
corresponding to $v_\text{tang}$, hence its shape $D\times(D-1)$. 
This eigenvector $v_\text{tang}$ provides the tangent of each image of
the instanton. The translational and rotational eigenvectors are
constructed as described elsewhere.\cite{wil80}

\begin{figure}%[htbp!]
\begin{center}
  \includegraphics[width=8.cm]{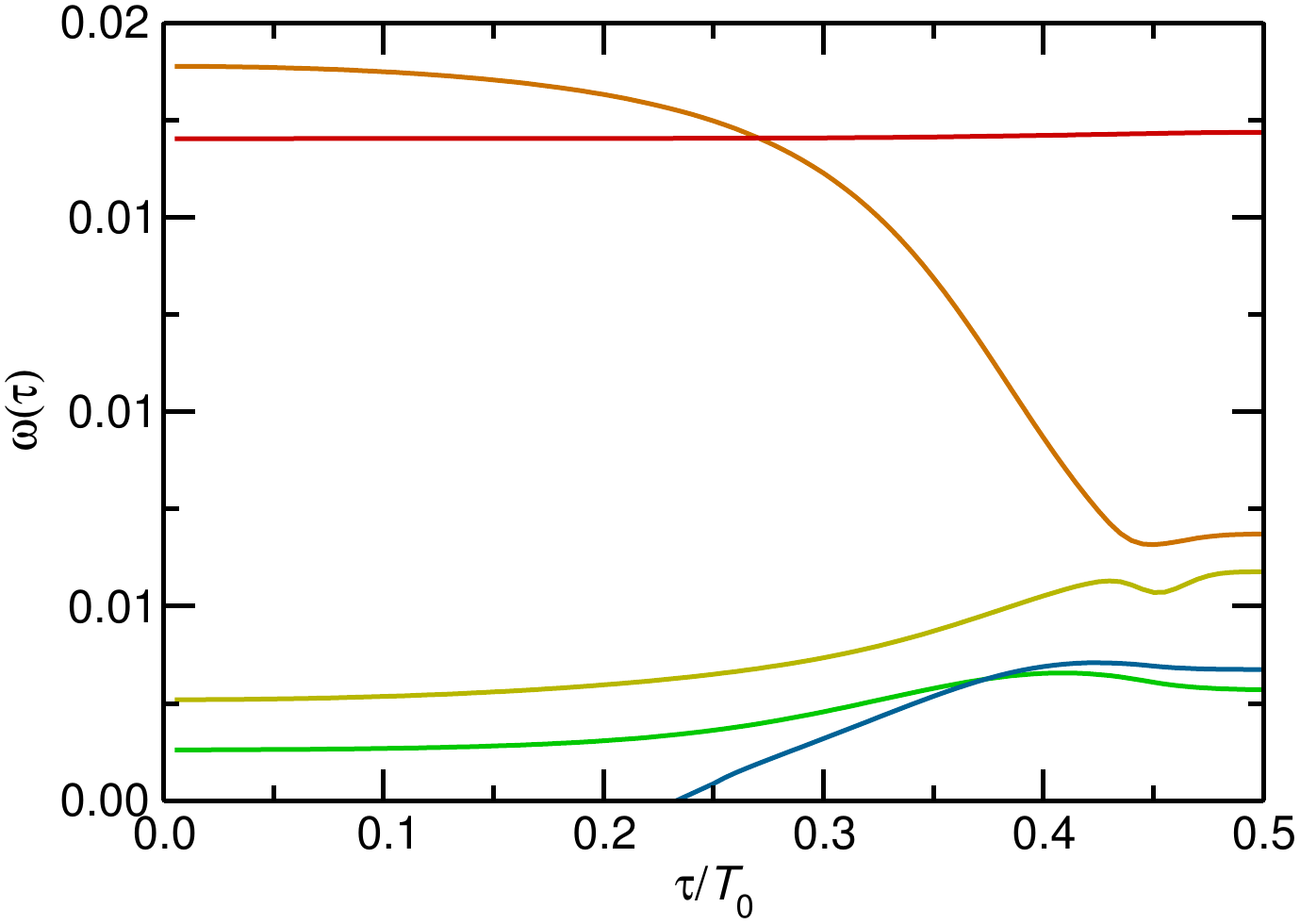}
    \caption{\label{fig:eval_trace} Vibrational frequencies of perpendicular modes traced along the instanton path for the reaction 
   H$_2$ + OH $\rightarrow$ H$_2$O + H on the
      NN1\cite{che13} potential energy surface. 
      % new graph at 200K
    }
\end{center}
\end{figure}

In order to average the eigenvectors to obtain \rem{$\omega_i(E)$}\add{$u_i(E)$},
equivalent modes need to be traced along the instanton path. Such a
tracing is possible if $\bfVtilde$ is constructed on a carefully
chosen coordinate system.  A Gram--Schmidt process is used to generate
$\bfyo$ at an arbitrary starting coordinate and to orthogonalize all
$D-1$ components of $\bfyo$ to the tangential vector and all unit
vectors of rotation and translation. An initial guess basis is
supplied to the Gram--Schmidt algorithm which can be arbitrarily
chosen for the first image.  The eigenvectors of $\bfVtilde$ are found
and saved in this reduced basis. For the neighboring image, the
process is repeated, new vectors tangential to the path and for
rotation and translation are found and a guess must be provided for
the remaining $D-1$ vectors, this time the set of $D-1$ saved eigenvectors
of the previous step having the smallest projection on to the 
instanton path are used as the new guess vectors. This ensures that the
Gram--Schmidt process produces a new coordinate system which is
similar to the coordinate system of the previous step.  The
eigenvectors of neighboring $\bfVtilde$ are represented in roughly
similar orthogonal bases, making their eigenvector comparison
possible. The maximum of the dot-products between eigenvectors of
successive images indicate the connection of the modes along the
instanton paths. This process is repeated for all images along the
path. The stability parameter $u_i(E)$ is then simply the
arithmetic average of the square roots of the eigenvalue of mode $i$.
Application of the eigenvalue tracing along the instanton path is shown in 
\figref{fig:eval_trace}. 

\addr{\Eqref{eq:uofomega} is only exact for separable systems and for
  a collapsed instanton. In non-separable systems it approximates 
  equations~(\ref{GYeq}) and (\ref{GYF}) quite well as it can be seen
  in \figref{fig:mic:stabpar}.}
\add{An obvious problem with eigenvalue tracing arises when too few images are 
used to localize the instanton path. If there is a loss of coherence between 
neighbouring $\bfVtilde$ then modes which should be identified as 
connected/distinct may be misclassified. The consequence of this can be seen in 
\figref{fig:eval_trace}, the where the crossings might 
switch to an avoided crossing.}

%--------------------------------------------------------------
\subsubsection{Frequency averaging to approximate $\sigma$\label{sec:freq_ave}}

Especially for a low number of images $P$, eigenvalue tracing becomes
numerically unstable. In order to approximate $\sigma$, however, it is
sufficient to know the trace of the square root of the Hessian matrix, which
will be justified in the following.  For large frequencies or large $T_0$,
$2\sinh(u_i/2)$ can be approximated by $\exp(u_i/2)$ which turns
\eqref{eq:sigma_def} into
%EQUATION 30
\begin{equation}
  \sigma(E)\approx\sum_{i=1}^{D-1} \frac{u_i(E)}{2}.
  \label{eq:mic:sinhexp}
\end{equation}

To obtain this, we consider the second line of \eqref{eq:sigma_def},
%\begin{align}\label{Kryv_action_ave_freq}
%  \sigma=&\int_{0}^{T_0}\bfyo(\tau)^T\cdot\left(\frac{1}{2}
%           \frac{d^2}{d\tau^2}+\frac{1}{2} 
%           \bfV(\bfyp(\tau))\right)\cdot\bfyo(\tau)d\tau
%\end{align}
and recognize the term in round brackets, ${\color{blue}-}\frac{1}{2}
\frac{d^2}{d\tau^2}+\frac{1}{2} \bfV(\tau)$, is the Hamiltonian for \rem{uncoupled,}
quantum harmonic oscillators. This can be \add{replaced with a diagonal}
\rem{diagonalized to obtain an}
energy-eigenvalue matrix \add{$\bomega(\tau)$} 
%EQUATION 31
\begin{align}\label{Kryv_action_QHO}
%  \sigma=&\mathrm{tr}\left( \int_{0}^{T_0}\bfyo(\tau)^T
%           \frac{\bomega(\tau)}{2}\bfyo(\tau)d\tau\right)
\bfyo^{(i)}(\tau)^T\left({\color{blue}-}\frac{1}{2}
           \frac{d^2}{d\tau^2}+\frac{1}{2} 
           \bfV(\tau)\right)\bfyo^{(i)}(\tau)\to&\frac{\bomega(\tau)_{ii}}{2}
\end{align}
\add{The matrix $\bomega(\tau)$ does not contain the eigenvalues of $\bfV(\tau)$, 
but rather the eigenvalues of the reduced Hessian $\bfVtilde(\tau)$.}		
In the limit $T_0\to\infty$, only the ground states of the \rem{uncoupled}
quantum harmonic oscillators make a significant contribution to the
integral in \eqref{eq:sigma_def}, thus the elements of $\bomega(\tau)$ are
\rem{$\bomega(\tau)_{ij}=\delta_{ij}\omega_{i}(\tau)$} \add{$\bomega(\tau)_{ij}=\delta_{ij}\tilde{\omega}_{i}(\tau)$}. Note, that \rem{$\omega_{i}$} \add{$\tilde{\omega}_{i}$}
are the energy-eigenvalues, i.e., the square roots of the eigenvalues of \rem{$\bfV(\tau)$} \add{$\bfVtilde(\tau)$}.
Using the replacement in \eqref{Kryv_action_QHO}, 
we discretize the integral over $d\tau$ in \eqref{eq:sigma_def} 
\rem{and project at all times $\tau$, onto the fixed basis $\bfyoo$; one may
  then rewrite \eqref{eq:sigma_def}} \add{which} may be rewritten
%EQUATION 32
\begin{align}\label{sigma_def_2}
  &e^{\sigma}={\prod_{i=1}^{D-1}\int d\bfyoo^{(i)}\bfyoo^{(i)T}\Bigg[e^{\sum_{j=1}^{P}\left(
              \frac{\bomega^{(j)}_{ii}}{2}\right)\Delta\tau}
              \Bigg]\bfyoo^{(i)},}\nonumber\\
  &e^{\sigma}=\prod_{i=1}^{D-1}\prod_{j=1}^{P}\int d\bfyoo^{(i)}\bfyoo^{(i)T}\Bigg[e^{\Delta\tau\frac{\bomega^{(j)}_{ii}}{2}}\Bigg]\bfyoo^{(i)}.
\end{align}
\rem{\Eqref{sigma_def_2} is exact in the limit $P\to\infty$, which
also implies $\bfP_i^T\bfP_{i+1}\to\mathbf{I}$.  Furthermore, by
definition, $\bfP_1=\bfP_P=\mathbf{I}$.}  
\rem{The fixed basis $\bfyoo$ is the perpendicular basis at the starting image. 
The integration is performed over a closed loop, hence the starting image may 
be chosen arbitrarily. Furthermore, by definition $\bfP_0=\bfP_P=\mathbf{I}$. In the limit of infinite images
$P\to\infty$ the following condition is true $\bfP_i^T\bfP_{i+1}\to\mathbf{I}$, 
this condition is assumed to be satisfied for any finite number of images.}
\add{where $\bfyoo$ is an arbitrary, fixed, perpendicular basis.}
We are left with the
definition of an operator trace. This formulation for $\sigma$ has
been used elsewhere.\cite{kry13}
%EQUATION 33
\begin{align}\label{sigma_derivation}
  e^{\sigma}=&\int d\bfyoo \mem{\bfyoo}
               {e^{\sum_{j=1}^{P}\Delta\tau
               \frac{\bomega^{(j)}}{2}}}{\bfyoo},\nonumber\\
  \sigma\approx&\frac{T_0}{2P}\sum_{j=1}^{P}
              \mathbf{tr}\left(\bomega^{(j)}\right)
\end{align}  
We see that \eqref{sigma_derivation} corresponds exactly to
\eqref{eq:mic:sinhexp} given the trace is conducted over
$D-1$ dimensions and if the $P$ images are spaced evenly in imaginary time.  

In practice, \rem{the projection operators} \add{the eigenvalues $\tilde{\omega}_i$ of $\bfVtilde$} don't need to be calculated
explicitly. The square roots $\omega_ {n,j}$ of the eigenvalues of the 
Hessian-matrices $\bfV_j$ of all images $j$ are calculated. For each image,
all $D$ of these $\omega_n(\tau)$ which belong to any vibrations are summed
up. In order to remove the contribution along the path, $\sigma$ is then
obtained as
%EQUATION 34
\begin{equation}
\sigma\approx \frac{T_0}{2P}\Re\left[\sum_{j=1}^{P} \sum_{n=1}^D \omega_{n,j} - \sqrt{\langle
v_{\text{tang},j} | \bfV_j | v_{\text{tang},j} \rangle}\right]
\end{equation}
with $v_{\text{tang},j}$, again, being the tangent of the instanton path at
image $j$. \add{It may be that $\omega_{n,j}$ is imaginary or $\langle
  v_{\text{tang},j} | \bfV_j | v_{\text{tang},j} \rangle$ is negative. In both
  cases, these contributions need to be ignored, only the real part is used.}
Since this expression avoids any eigenvalue tracing and only requires the sum
of the eigenvalues at each image it is numerically more stable for few images
or low energies. It can be expected to be accurate for large
\rem{$\omega_i(E)$} \add{frequencies}
and/or large $T_0$, i.e., when the approximation in \eqref{eq:mic:sinhexp} is
valid.

%--------------------------------------------------------------
\subsubsection{The product of eigenvalues of the full Hessian to approximate
  $\sigma$} 

The numerically most stable fall-back option we found is to use the Hessian of
the full instanton, i.e., the matrix of second derivatives of the Euclidean
action with respect to all atom coordinates of all images. This matrix is
required to calculate the temperature-dependent rate constant directly
(``canonical instanton'').\cite{rom11b} The eigenvalues $\lambda_i$ of that full
Hessian accounts for the fluctuations perpendicular and along the instanton
path. It is not directly possible to obtain $u_i$ from these, but since
$\sigma$ covers the fluctuations perpendicular to the path, this is available
by projection.

The eigenvectors associated to the $\lambda_i$ are denoted by
$v_i$. One $\lambda$ is zero, its eigenvector $v_\text{tang}$ provides
the tangent to the instanton path. It is scaled to have unit-length
for each image.  $\langle v_i|v_\text{tang}\rangle$ is the projection
of an arbitrary eigenvector on the tangent. It is between between 0
and 1 for each eigenvector,
$\sum_{i=1}^{NP-1}\langle v_i|v_\text{tang}\rangle^2=P-1$. With that,
the fluctuations perpendicular to the path result in
%EQUATION 35
\begin{equation}
  \exp(\sigma)=
  \left(\frac{\beta}{P}\right)^{(N-1)P}\ 
  \prod_{i=1}^{NP-1}\sqrt{|\lambda_{\text{inst},i}|^{1-\langle v_i|v_\text{tang}\rangle^2}}
  \label{eq:qts:kry6}
\end{equation}
where $N$ is the number of degrees of freedom in the system, not the number of
atoms. \Eqref{eq:qts:kry6} is correct if there are no zero modes due to
rotation and translation. If there are, however, they have to be taken out of
the product, this could be done via similar projections as in
\eqref{eq:qts:kry6}. However, we use the analytical expression of the
eigenvalues of the instanton-Hessian relating to zero vibrational frequencies:
%EQUATION 36
\begin{equation}
  \lambda_{0,i}=4\left(\frac{P}{\beta}\right)^2\sin^2(i\pi/P)
  \quad i=1,\ldots,P
\end{equation}
The product of all these eigenvalues except the last one which is zero
is termed $A_0$:
%EQUATION 37
\begin{equation}
  A_0=\prod_{i=1}^{P-1}\lambda_{0,i}
\end{equation}
with that, for $N_0$ zero modes ($N_0=N-D$)
%EQUATION 38
\begin{multline}
  \exp(\sigma)=
  \left(\frac{\beta}{P}\right)^{(N-1-N_0)P} \frac{1}{A_0^{N_0}}\times\\
  \prod_{i=1}^{NP-1-N_0}\sqrt{|\lambda_{\text{inst},i}|^{1-\langle {v_i|v_\text{tang}}\rangle^2}}
  \label{eq:qts:kry6a}
\end{multline}
where all zero eigenvalues are ignored in the product. From that, we
obtain $\sigma'$ as $\sigma'=\sigma/T_0$. This is only accurate if
\eqref{eq:mic:sinhexp} is fulfilled, i.e. for large frequencies.

Apart from the methods described here, which turned out to be the most
promising ones, we tested several other approaches. For example rather
than averaging frequencies, it is possible to average the Hessian
matrices $\bfV$ with the component tangentially to the path
removed. The eigenvalues of the resulting averaged Hessian can be used
as approximations for \rem{$\omega_i(E)$}\add{$u_i(E)/T_0(E)$}. For an instanton path which couples
strongly to other vibrational modes, the removal is not exact, though.

\begin{figure*}%[htbp]
\begin{center}
  \includegraphics[width=8cm]{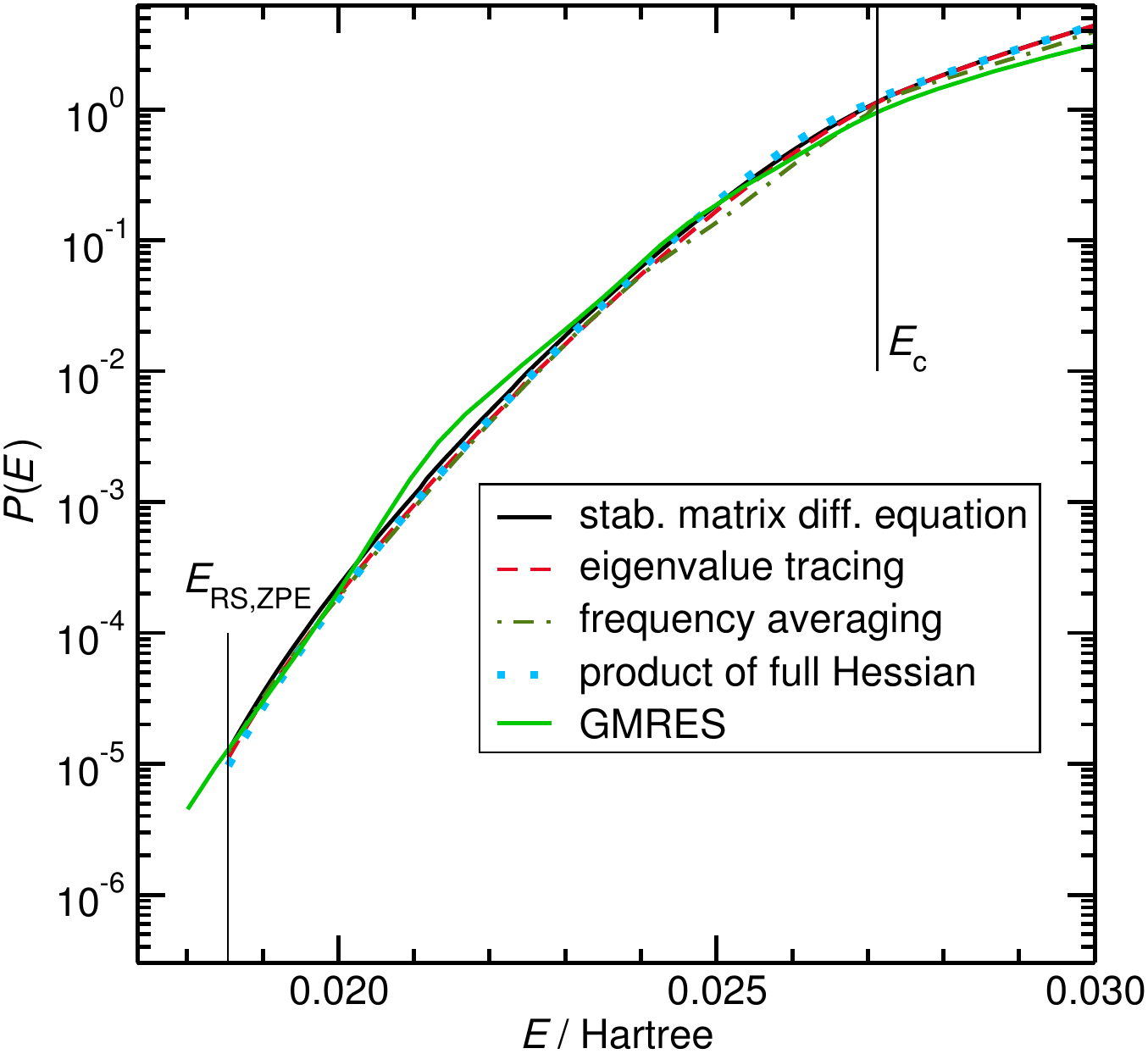}
  \includegraphics[width=8.2cm]{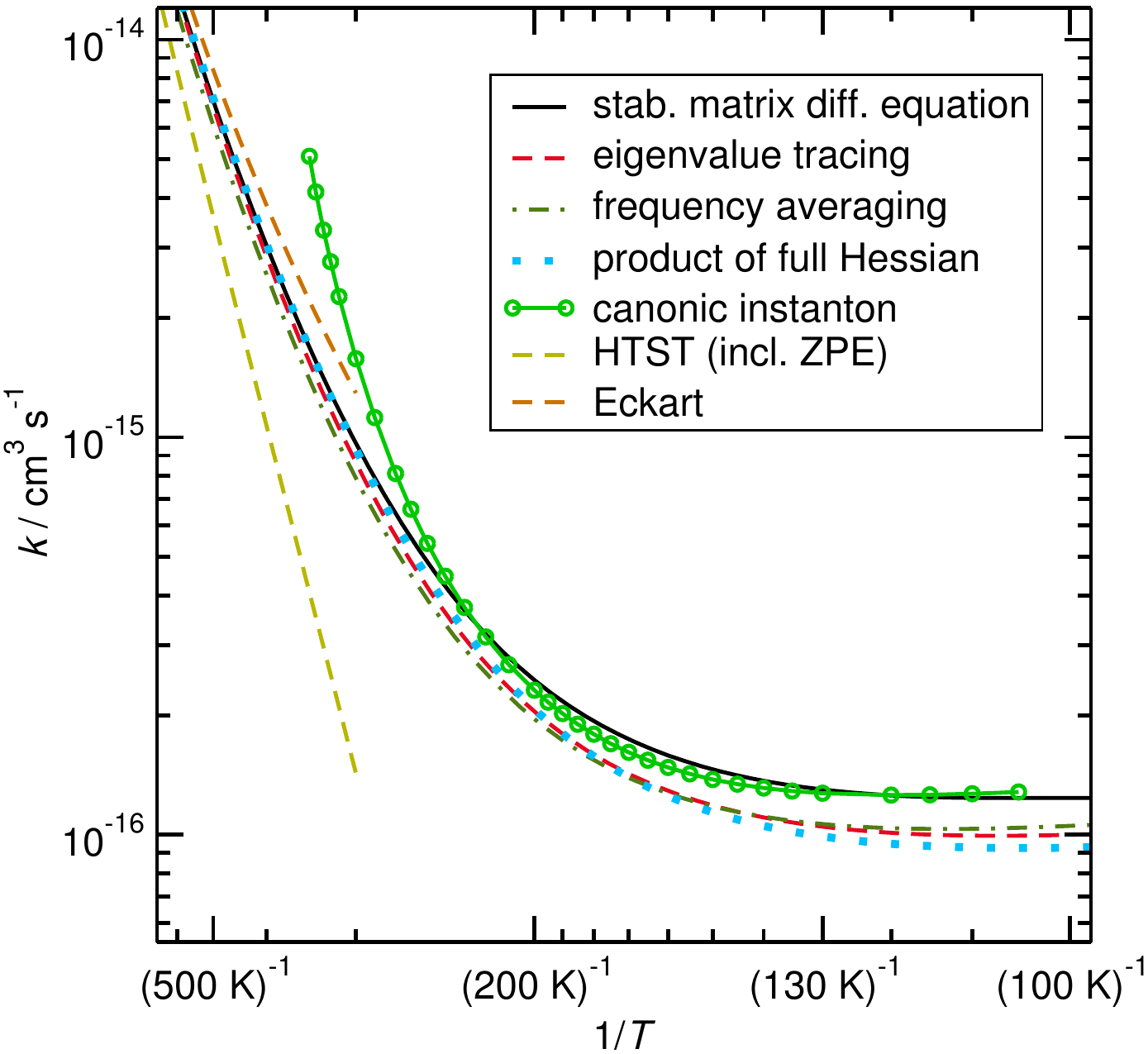}
  \caption{$P(E)$ for the reaction H$_2$+OH on the Schatz--Elgersma PES
    \cite{wal80,sch80} for the different approximations of $\sigma'$ or
    $u_i(E)$ discussed in the methods section. GMRES refers to
    \cite{man93,man94} and acts as a reference. Thermal rate constants
    are shown in the left panel.
    \label{fig:h3oschatz}
  }
\end{center}
\end{figure*}

%%%%%%%%%%%%%%%%%%%%%%%%%%%%%%%%%%%%%%%%%%%%%%%%%%%%%%%%%%%%%%%%%%%%%%%%
\section{Applications}
%%%%%%%%%%%%%%%%%%%%%%%%%%%%%%%%%%%%%%%%%%%%%%%%%%%%%%%%%%%%%%%%%%%%%%%%

As a numerical test of our derivations we apply them to two chemical systems,
one described by a fitted potential energy surface, the other one with
energies, gradients and Hessians calculated on the fly from DFT. As a first
test system we chose the reaction H$_2$ + OH $\rightarrow$ H$_2$O + H, which
has been investigated in great detail in the
literature.\cite{man93,man94,zha94, mil98a, man00,yan01a,cha04, bha10,
  fu10,bha11, esp10, ngu11,che13,fu15} Among the many potential energy
surfaces (PES) available, we use the old one by Schatz and
Elgersma.\cite{wal80,sch80} The reason is that for this surface, we have
``exact'' reference data for $P(E)$ from quantum dynamics calculations with a time-independent grid
representation and the generalized minimum residual (GMRES) method available.\cite{man93,man94} 
The choice of the potential and the system was
made to compare to other methods rather than to provide new physical insight
into that particular reaction.  The results we obtained are shown in
\figref{fig:h3oschatz}. All calculations were done in
DL-FIND.\cite{kae09a}

The results for $P(E)$ are shown in the left panel of
\figref{fig:h3oschatz}. While it is possible to find instantons for
$E<E_\text{RS,ZPE}$, the resulting data are not necessary for the calculation
of thermal rate constants (right panel) using \eqref{eq:mic:can:final}. All
methods we tested show a similar dependence of $P$ on $E$. This four-atom
system has five stability parameters $u_i(E)$ perpendicular to the instanton
path. Since here we have an analytical expression for the potential energy
surface, we can use many images (up to $P=400$ was used), so that even solving
the stability matrix differential equation \eqref{GYeq} is stable enough. The
results are very similar to eigenvalue tracing. In both of these approaches,
\eqref{eq:mic:stabpar} and (\ref{eq:mic:2}) were used to obtain $P(E)$. Even
though $P(E)$ obtained from solving the stability matrix differential equation
and eigenvalue tracing look very similar, the resulting thermal rate constants
are somewhat different ($\sim$13\%) at low temperature. While this emphasizes
that very accurate cumulative reaction probabilities are required to calculate
thermal rate constants in the range of deep tunneling, such small differences
are probably negligible in practical applications. Obtaining $P(E)$ from
\eqref{eq:mic:psigma} via $\sigma$ using frequency averaging or obtaining
$\sigma$ from the full Hessian of the Euclidean action results in pretty good
approximations as well. \rem{Frequency averaging consistently underestimates $P(E)$
and, consequently, $k(T)$. Using the full Hessian gives very consistent rate
constants to the two methods which provide individual $u_i$, however.} This is
the case even though rather small frequencies perpendicular to the instanton
are present, in which case the approximation of \eqref{eq:mic:sinhexp} may be
questioned. It seems to work well in practice, though. The smallest frequency
at the reactant side of the instanton (in the pre-reactive minimum) is only
173~cm$^{-1}$ ($\omega=7.9\times 10^{-4}$ a.u.) on the PES we used.

\begin{figure*}%[htbp!]
\begin{center}
  \includegraphics[width=8cm]{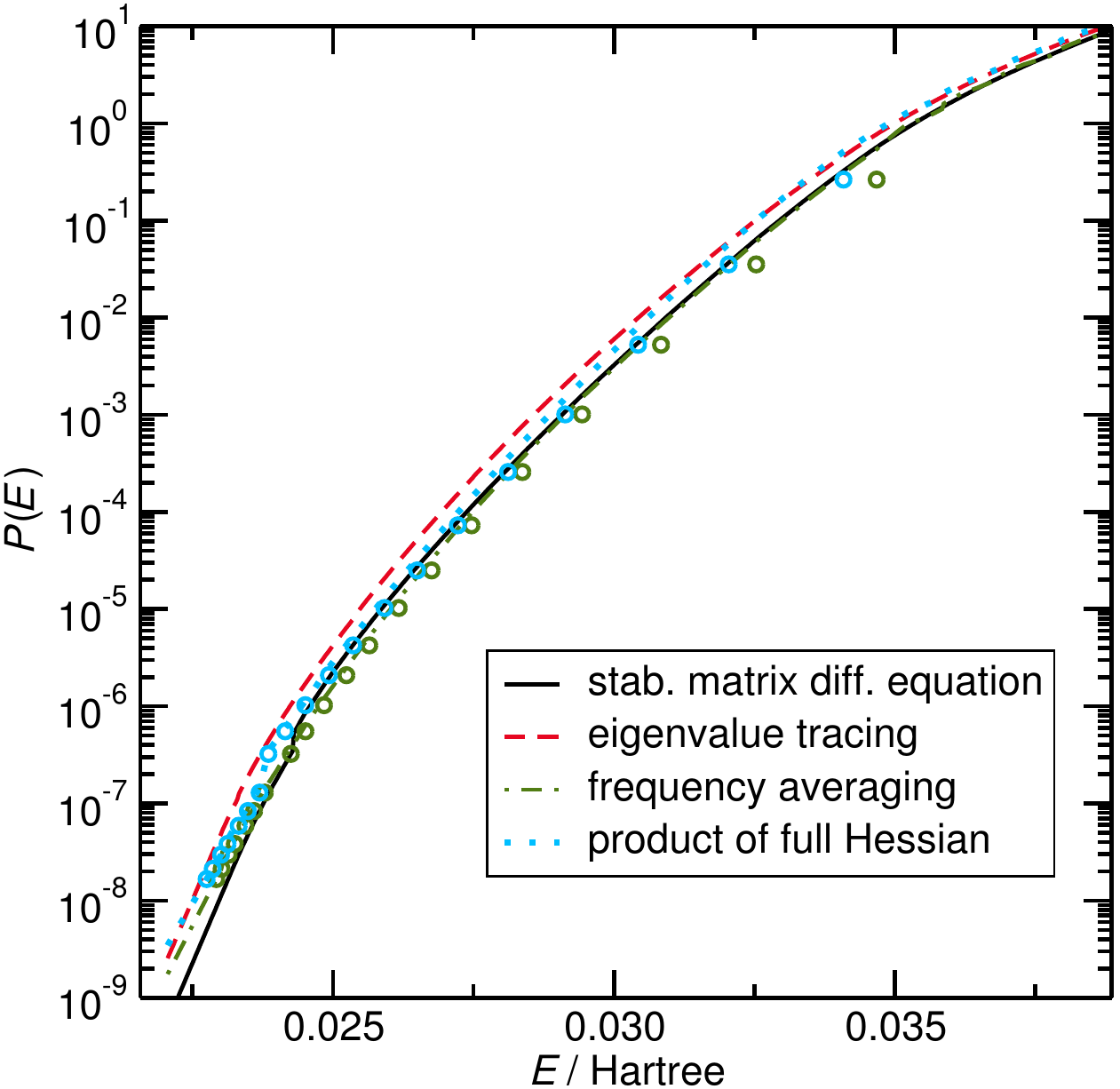}
  \includegraphics[width=8cm]{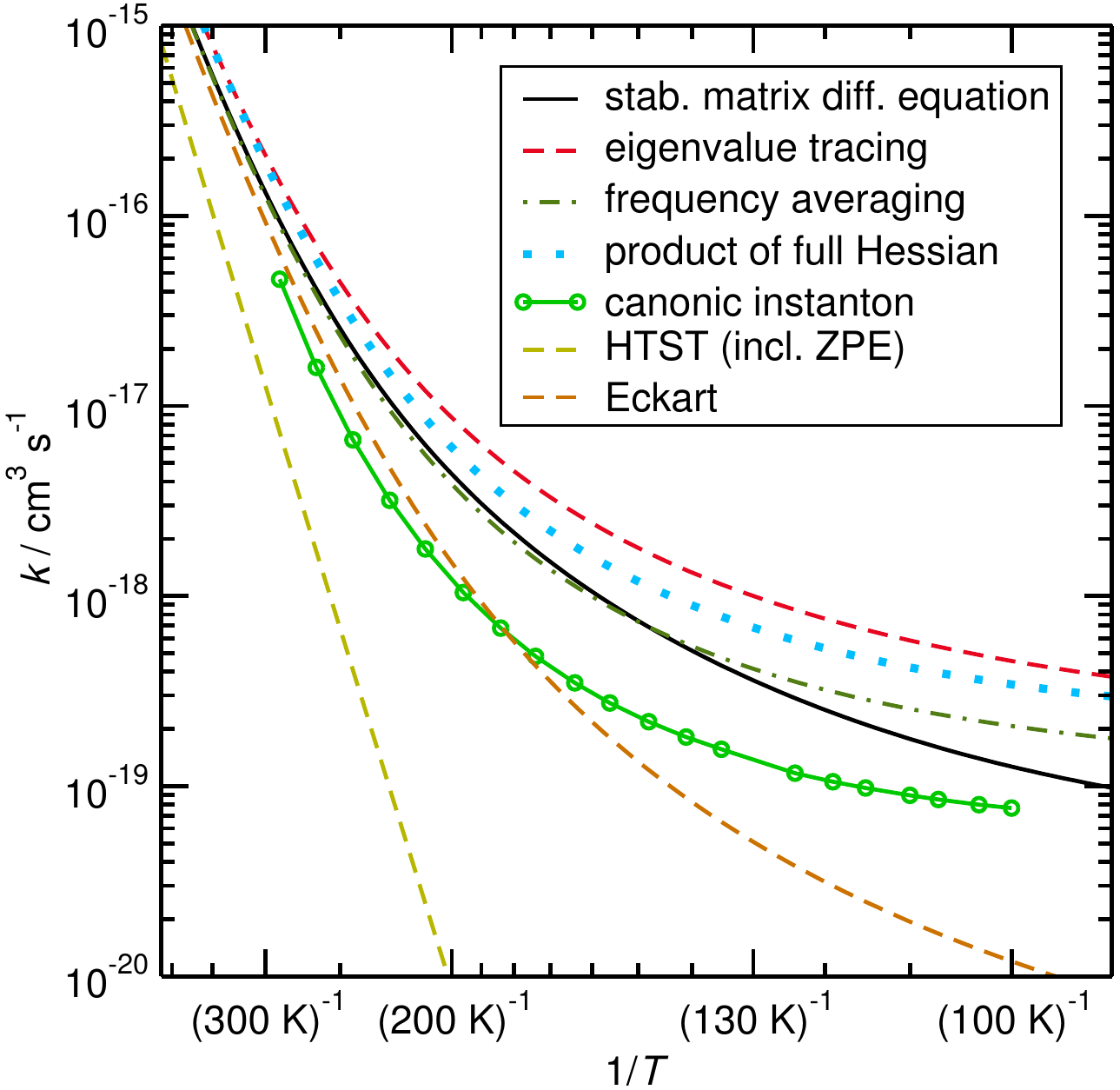}
  \caption{$P(E)$ and thermal rate constants
    for the reaction HNCO + H
    $\rightarrow$ NH$_2$CO for the different approximations
    of $\sigma'$ or $u_i(E)$. In the left graph, open circles show
    $[1+\exp(S_0)]^{-1}$ at $E=E_b+\sigma/T_0$ for each instanton, cf. \eqref{eq:mic:psigma}.
    \label{fig:hnco}
  }
\end{center}
\end{figure*}

For comparison \figref{fig:h3oschatz} and \figref{fig:hnco} also show
rate constants calculated with harmonic transition state theory (HTST)
with the full quantum mechanical partition functions of harmonic
oscillators used for all vibrations. Thus, they include ZPE, but no
tunneling. Additionally, a curve \add{with HTST corrected for}
\rem{with} tunneling \rem{approximated} \rem{by} \add{through} a
symmetric Eckart barrier (height and $\bar\omega_{\text{TS}}$ matched
to the PES) \rem{additional to HTST} is shown.

\begin{figure}%[htbp!]
\begin{center}
  \includegraphics[width=8cm]{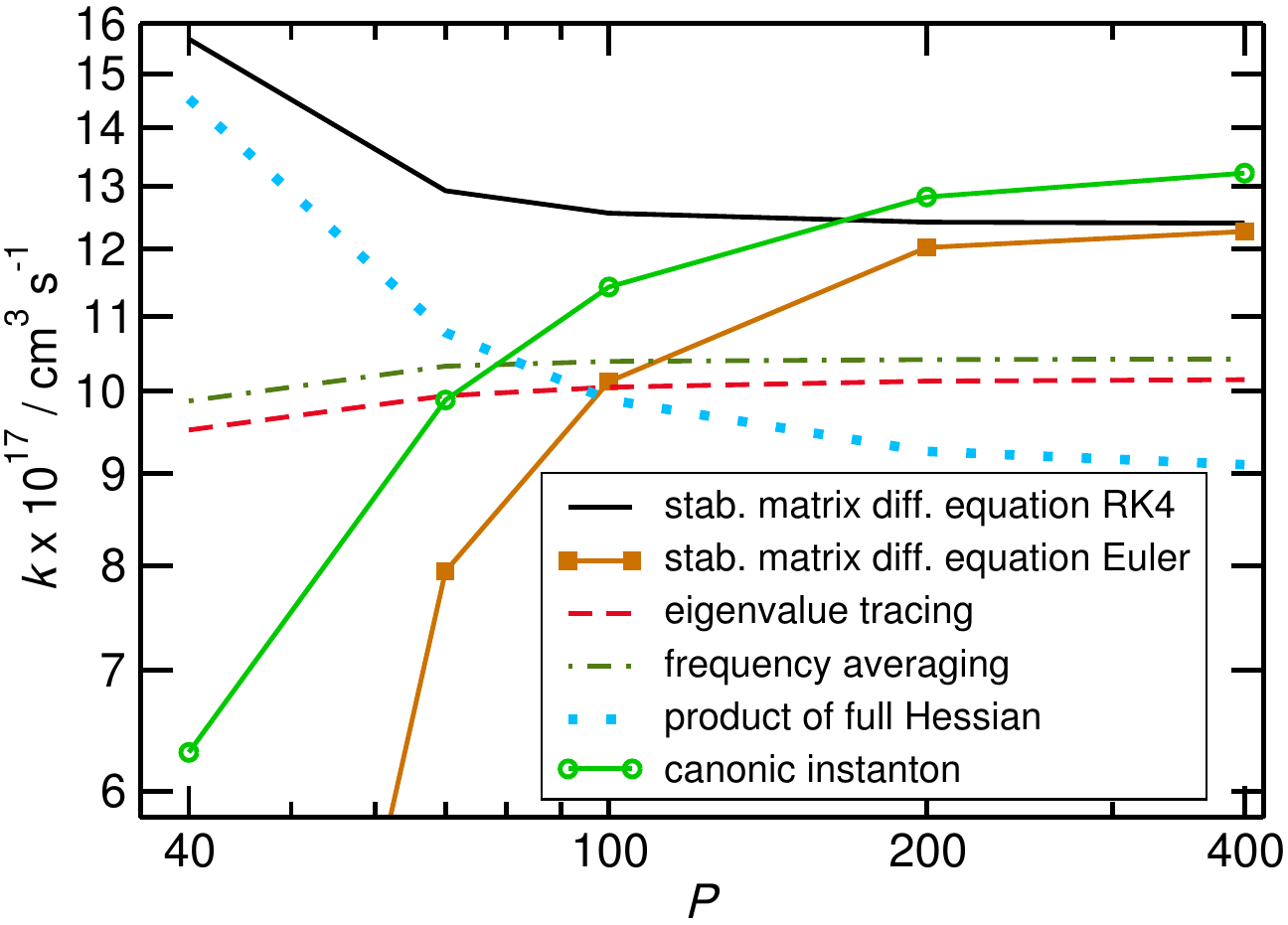}
  \caption{Robustness of the different methods to calculate the rate constant
    with change in the number of images $P$ of the instanton. Rate constants
    at $T=105$ K for the reaction H$_2$ + OH $\rightarrow$ H$_2$O + H on the
    Schatz--Elgersma surface\cite{wal80,sch80} are compared. \add{The solution
      of the stability matrix differential equation was performed with a 
      fourth-order Runge--Kutta method (RK4), as well as with the backward
      Euler approach.}
    \label{fig:mic:nimage}
  }
\end{center}
\end{figure}

Our values for the cumulative reaction probability $P(E)$ agree very
well with the reference values found by quantum dynamics. Small
fluctuations are smoothed out by the thermal averaging. This means
that the approximations made in instanton theory, most of all the
semiclassical approximation, are valid for this system.

A large number of images $P$ can be used for a small system like H$_2$ + OH
$\rightarrow$ H$_2$O + H, especially when using a fitted potential energy
surface. When the energies, gradients, and Hessians must be calculated on the
fly by electronic structure theory, however, $P$ is limited. Therefore, we
checked how strongly the thermal rate constant obtained via
\eqref{eq:mic:can:final} with $P(E)$ calculated with the different methods
depends on $P$. The results for one temperature ($T=105$ K) are shown
in \figref{fig:mic:nimage}, comparison for the full temperature range
is given in the supporting information. The number
of images $P$ has been kept constant for the whole range of $E_b$. It is
obvious from \figref{fig:mic:nimage} that frequency averaging and eigenvalue
tracing are rather stable at few images while $\sigma$ obtained from the
product of the full Hessian causes a significant error for small
$P$. The solution of the stability matrix differential equation is also
\add{somewhat} sensitive to the number of images and breaks
down for $P=40$ \add{when using the Euler method}. It should be noted that canonical instanton results at the same temperature also
depend strongly on $P$, which is well-known.\cite{and09,rom11b} 

\begin{figure}%[h!]
\begin{center}
  \includegraphics[width=8cm]{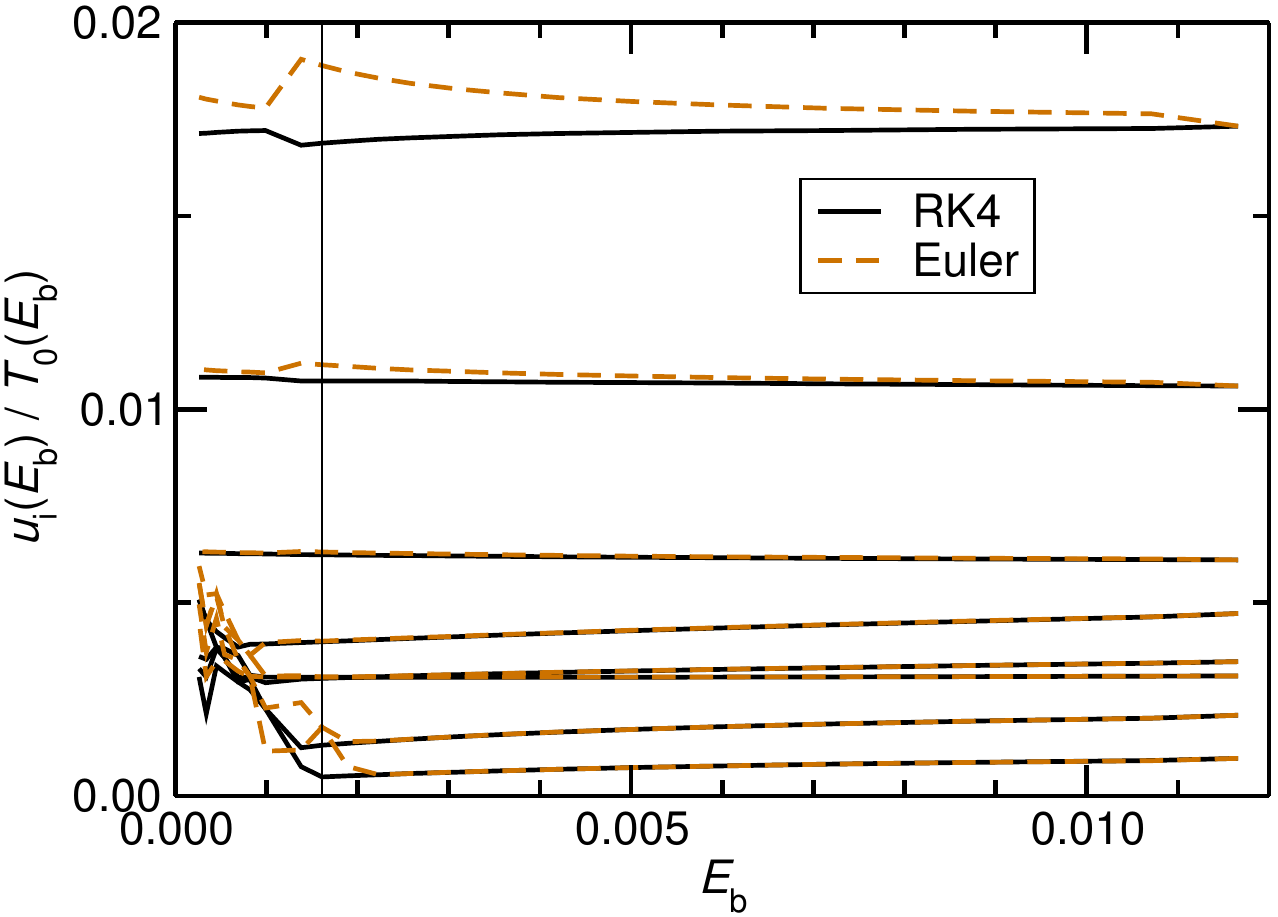}
  \caption{\add{The stability parameters for the reaction HNCO +
      H $\rightarrow$ NH$_2$CO obtained by solving the stability matrix
      differential equation by a fourth-order Runge--Kutta approach and by the
    backward Euler algorithm. Both are unstable at lower $E_b$ than the thin
    vertical line.}
    \label{fig:mic:stabpar:hnco}
  }
\end{center}
\end{figure}

In order to compare our approaches on a yet more realistic calculation, we
applied them to the reaction H + HNCO $\rightarrow$ NH$_2$CO for which
energies, gradients and Hessians were obtained on the fly from density
functional calculations. The new calculations were done in
DL-FIND\cite{kae09a} via ChemShell,\cite{she03,met14} details of the
theoretical treatment are given elsewhere.\cite{son16} The energy and its
derivatives contain numerical noise due to the incompleteness of the SCF
iterations and other approximations. Only $P=40$ images were used to optimize
instantons down to 135~K and $P=78$ below that. Instantons down to 100~K were
used, which is not quite sufficient to obtain $P(E)$ down to
$E_\text{RS,ZPE}$. The effect can be seen in \figref{fig:hnco}: at low
energies, $P(E)$ had to be extrapolated. The change in $P$ leads to a
noticeable step in $P(E)$. Stable solution of the stability matrix
differential equation \rem{was impossible for all energies} \add{could only be
  achieved for high energies, see \figref{fig:mic:stabpar:hnco}. The limit of
  stability is indicated as a thin vertical line. The last value of
  $u_i(E)$ to the right of the line was used at lower $E$}. The canonical rate
constants agree reasonably well with those obtained from canonical instanton
theory. It is obvious from \figref{fig:hnco} that \rem{eigenvalue tracing and
obtaining $\sigma$ from the full Hessian overestimate the rate constant
compared to} \add{all methods lead to higher rate constants than} canonical instanton theory. \rem{Frequency averaging results in a good
agreement with canonical instanton theory, but this may be purely coincidental.} 
Overall, it is clear that real-world applications with numerical
noise in the potential energy surface and limited $P$ lead to challenges but
can be successfully treated with the algorithms presented.

%%%%%%%%%%%%%%%%%%%%%%%%%%%%%%%%%%%%%%%%%%%%%%%%%%%%%%%%%%%%%%%%%%%%%%%%
\section{Discussion}
%%%%%%%%%%%%%%%%%%%%%%%%%%%%%%%%%%%%%%%%%%%%%%%%%%%%%%%%%%%%%%%%%%%%%%%%

Microcanonical instanton theory allows the calculation of rate constants in
bimolecular reactions under low-pressure conditions, i.e. in conditions in
which a pre-reactive minimum will not thermally equilibrate. It also provides
thermal rate constants over the full temperature range. The problem of canonical
semiclassical instanton theory breaking down at $T_\text{c}$ is avoided
intrinsically. Both of these advantages offer significant opportunities to
study chemical processes at low temperature. 

The use of microcanonical instanton theory poses challenges as well, however. In
order to calculate any thermal rate constant, instanton calculations along the
whole temperature or energy range ($T_0$ corresponds to
$\beta\hbar=\hbar/(k_\text{B}T)$ in canonical calculations) from $T_\text{c}$
to $E_b+\sigma/T_0<E_\text{RS,ZPE}$ need to be performed in principle. In
practice, one can extrapolate to some extent, as shown for the case of H + HNCO
$\rightarrow$ NH$_2$CO in \figref{fig:hnco}. Each of these instanton
calculations needs to be converged with respect to the number of images
$P$. Convergence can be slow, as shown in \figref{fig:mic:nimage}, but is
still generally faster than for canonical instanton theory. In the latter, the
rate constant at a specific temperature, as long as this temperature is well
below $T_\text{c}$, can be provided with high accuracy by converging with
respect to $P$. Using microcanonical theory, it is easier to provide a rough
approximation of the rate constant over a larger temperature range. 

Another important difference between canonical and microcanonical instanton
theory is the number of choices of methods an approximations. In canonical
instanton theory, over the last years a set of reliable algorithms was
established: searching for instantons using a modified Newton--Raphson
converges fast and reliably,\cite{rom11b} the rate constant is calculated via
the diagonalization of the full Hessian of the Euclidean
action\cite{arn07,rom11} and the rotational partition function, using
$J$-shifting, is calculated from the geometries of the images\cite{and09} along the
instanton path.\footnote{All these are based on earlier work, but
  described nicely in the references mentioned here.} The only remaining choice
or parameter is the number of images. Rate constants need to be converged with
respect to that. In the microcanonical case, such a generally recommendable
algorithm is not established yet. Not even a unique, recommendable rate
expression is known at present. \Eqref{eq:mic:noshift} is the direct
consequence of the semiclassical treatment of \eqref{final}, but inapplicable
in practice. The combination of \eqref{eq:mic:stabpar} and \eqref{eq:mic:2}
seems a promising way out, but using \eqref{eq:mic:psigma} may be equally
justified. Even with a given rate expression, the question of how to calculate
the stability parameters $u_i(E)$ \rem{or $\omega_i(E)$} or their combination
$\sigma$ remains. Here, we presented the four most promising approaches of
several others we have tried. Practical application to many other cases will
have to show which algorithm proves the most promising.

%%%%%%%%%%%%%%%%%%%%%%%%%%%%%%%%%%%%%%%%%%%%%%%%%%%%%%%%%%%%%%%%%%%%%%%%
\section{Conclusions}
%%%%%%%%%%%%%%%%%%%%%%%%%%%%%%%%%%%%%%%%%%%%%%%%%%%%%%%%%%%%%%%%%%%%%%%%

Based on different rate expressions for microcanonical instanton rate
constants, we have proposed and tested several algorithms to calculate the
required stability parameters for non-separable molecular systems,
i.e. systems in which the vibrational modes perpendicular to the instanton
path couple to vibrations along the path. Any realistic molecules are
non-separable. We found the traditional way of integrating the stability
matrix differential equation numerically unstable in general. Stability
parameters can, however, also be obtained as vibrational frequencies, averaged
along the instanton path. With that, the tracing of Hessian eigenvalues along
the instanton path and averaging of the corresponding frequencies leads to an
accurate and generally more stable algorithm to derive $u_i(E)$. An
alternative is to average all frequencies and use $\sigma$ rather than the
individual $u_i(E)$ in the rate expression. In that case, no tracing is
necessary, since all Hessian eigenvalues are averaged. Yet another approach is
to obtain $\sigma$ from all fluctuations of all images perpendicular to the
path by using the full Hessian of the Euclidean action. All methods presented
here have their merits in practical applications. Eigenvalue tracing and
frequency averaging were shown to be particularly stable at small numbers of
images to discretize the instanton path.

Overall, we provide viable approaches to calculate microcanonical instanton rate
constants and cumulative reaction probabilities. These have the advantage that
they provide rate constants over the whole temperature range without the
breakdown of canonical instanton theory at $T_\text{c}$ and its inaccuracies
close to it.

%\clearpage\newpage
\section{Supplementary Material}

Graphs showing the dependence of the rate constants on the number
of images for the different approximations.

\begin{acknowledgments} We thank Prof. Uwe Manthe for providing the original
  values of $P(E)$ for the quantum dynamics results reported a long time
  ago.\cite{man93,man94} This work was financially supported by the European
  Union's Horizon 2020 research and innovation programme (grant agreement
  No. 646717, TUNNELCHEM). AL received financial support by the Carl-Zeiss
  foundation.
\end{acknowledgments}

%\bibstyle{alpha}
%\bibliography{jabref}
%\bibliography{mod,loehle,Q_Chem_Uni_St}

%\bibliography{jabref,Q_Chem_Uni_St}
\bibliography{jabref}

% URL and PMID removed via makefile: rah03
% \bibliography{mod}

\end{document}